\documentstyle[12pt,aasms4]{article}
\def\puncspace{\ifmmode\,\else{\ifcat.\C{\if.\C\else\if,\C\else\if?\C\else%
\if:\C\else\if;\C\else\if-\C\else\if)\C\else\if/\C\else\if]\C\else\if'\C%
\else\space\fi\fi\fi\fi\fi\fi\fi\fi\fi\fi}%
\else\if\empty\C\else\if\space\C\else\space\fi\fi\fi}\fi}
\def\SP{\let\\=\empty\futurelet\C\puncspace }
\def\bj{b$_J$\SP }
\def\rf{r$_F$\SP }
\def\etal{et\SP al.\SP }
\def\micra{$\mu$m\SP }
\def\deg{$^\circ$\ }
\def\kms{kms$^{-1}$}

\begin{document}

\title{A Medium-Deep Survey of a Mini-Slice at the North
Galactic Pole. II. The Data\footnote{UCO/Lick Observatory Bulletin 1326}}
\author{C.N.A. Willmer\altaffilmark{2,3,4}}
\affil{CNRS, Institut d'Astrophysique de Paris, 98 bis, Boulevard
Arago, F-75014, Paris, France.}

\author{David C. Koo\altaffilmark{2} and Nancy Ellman\altaffilmark{5}}
\affil{UCO/Lick Observatory and Board of Studies in Astronomy and
Astrophysics, University of California, Santa Cruz, CA 95064.}

\author {Michael J. Kurtz}
\affil{Harvard-Smithsonian Center for Astrophysics, 60 Garden St.,
Cambridge, MA 02138.}

\and

\author{Alex S. Szalay\altaffilmark{2}}
\affil{Department of Physics, Johns Hopkins University, Baltimore, MD
21218.}

\altaffiltext{2}{Visiting Astronomer, Kitt Peak National Observatory.
KPNO is operated by AURA, Inc.\ under contract to the National Science
Foundation.} 

\altaffiltext{3}{Observat\'orio Nacional, Rua General Jos\'e Cristino
77, Rio de Janeiro, RJ 20921-030, Brazil.}

\altaffiltext{4}{UCO/ Lick Observatory, University of California,
Santa Cruz, CA 95064.}

\altaffiltext{5}{Princeton University Observatory,
Peyton Hall, Princeton NJ 08544.}

\begin{abstract}
We report 328 redshifts, \bj magnitudes and \bj--\rf colors of
galaxies measured in a redshift survey of a 4\deg\SP
$\times$ 0.67\deg\SP slice close to the north Galactic pole. The faintest
galaxies in this survey have a magnitude of b$_J$ $\sim$ 20.5. The
redshifts present external errors of the order of 70 \kms, and we
estimate that the mean photometry errors are $\sim$ 0.2 for magnitudes and
$\sim$ 0.3 for colors. The redshift completeness level of the sample
is of the order of $\sim$ 35\% at \bj=20, and part of this rather low
completeness is the result of the combination of limitations imposed by the
multifiber system with the clustering of galaxies, and an insufficient
number of configurations. At the nominal magnitude limit of the
survey, we were able to measure redshifts for $\sim$ 70 \% of the galaxies we
observed. From the correlation between observed properties of the
galaxies in this sample, we demonstrate that the mean surface
brightness is a major limiting factor in our ability to measure
redshifts of faint objects.
\end{abstract}
\clearpage

\section{Introduction}

The detection by Broadhurst et al. (1990) of an apparent repetition
pattern in the redshift distribution of galaxies toward both galactic
poles suggested that structures similar to the walls of galaxies
detected by the fully sampled wide-angle surveys of bright galaxies
($e.g.$, da Costa et al. 1988; Geller \& Huchra 1989) were common
occurrences. Further evidence supporting
this interpretation was presented by Koo et al. (1993), who used
several pencil-beam surveys separated by a few degrees close to the
direction of the galactic poles demonstrating
that the original peaks could not be due to rich groups or clusters of
galaxies, while in other directions of the sky similar patterns were
also detected. Although the sparsely-sampled pencil beam surveys are an
efficient means of probing the galaxy distribution on very large
scales, in order to have a better characterization of the properties
of these structures, which have a rather low projected surface
density, a higher sampling rate than used in the pencil-beam surveys
is necessary, as only a few galaxies per wall would be detected by the
original probes.
In a previous paper (Willmer et al. 1994, hereafter paper I) we
presented the results of a redshift survey for a 4\deg\SP $\times$
0.67\deg\SP mini-slice which was designed to characterize the
properties of the four or five nearest peaks detected 
by Broadhurst et al. (1990) in the direction of the north Galactic
pole.  
This work took advantage of the availability in spring 1992 of the
HYDRA  multi-fiber spectrograph (Barden et al. 1992) which would allow
gathering up to 97 spectra simultaneously, thus allowing a denser
sampling rate compared to that previously available. The opening angle
of this survey was designed to cover a projected distance of $\sim$ 20
h$^{-1}$ Mpc at z $\sim$ 0.1 ($h$ = H$_0$/100 kms$^{-1}$ Mpc and
q$_0$=0.5 assumed throughout), which is the smallest possible
projected distance that allows a proper characterization of a
structure as a wall of galaxies according to the Monte Carlo
simulations carried out by de Lapparent, Geller \& Huchra (1991) and
Ramella, Geller \& Huchra (1992). Both works used the Great Wall as a
typical representative of walls of galaxies.

In this paper we report redshifts and magnitudes of galaxies that were
observed in this survey, as well as a discussion of the possible
selection effects that are present in this sample. The acquisition of
the catalog of galaxies and its photometric calibration is described
in section 2,  followed by the description of the  spectroscopic data
in section 3. The catalog and some of its properties are presented in
section 4. Some concluding remarks follow in section 5.

\section{Astrometry and Photometry}
\subsection{Detection of Objects}
As described in paper I, the sample used in this survey was derived
from scans of two recent epoch (1989, 1990) Palomar
Schmidt plates, SF02410 (IIIaF), and SJ03220 (IIIaJ). These plates had been
obtained for the POSS II (Reid et al. 1991), but because of image
elongation, they were eventually replaced by higher quality material.
 
The KPNO PDS microdensitometer was used in density mode to digitize
the F plate, using pixels of 10 \micra sampled at 10 \micra spacing,
each pixel corresponding to $\sim$ 0.6$''$ on the sky.
An area of about 4\deg\SP $\times$ 1\deg\SP centered at $\delta$=29.5\deg\SP
(Epoch 1950) was covered by thirty-five scans of 2000 $\times$ 2000 pixels,
with 200 pixels overlapping both in X and Y in order to
ensure a uniform magnitude scale 
and to preserve a global coordinate system. The densities 
were converted into intensities using the sensitometry spots and the
log (exposure) values given by Reid et al. (1991). The fit and transformation
were made using the DTOI
package of IRAF\footnote{IRAF is distributed by National Optical
Astronomy Observatories. NOAO is operated by AURA Inc. under contract
to the National Science Foundation}. The detection and classification
of 
images was carried out using FOCAS (Jarvis \& Tyson 1981; Valdes
1982). Objects were defined as groups of more than 50 connected pixels
($\sim$ 23 arc sec$^2$)
at a threshold of 3 $\sigma$ above the sky. From the catalog of objects
generated by FOCAS we selected the following parameters for each
object: the intensity-weighted X and Y positions; the isophotal
magnitude; the object area; the second inverse moment of the
light distribution (Kron 1980); and the automatic classification made
by the FOCAS algorithm (stars, galaxies, diffuse, long or noise) which
takes into account the shape of the stellar point spread
function. The latter was calculated using unsaturated stellar objects
detected in each scan. 

The  brightest stars of the Hubble Space Telescope
Guide Star Catalog (Lasker et al. 1990, hereafter referred to as GSC)
were identified visually and used to define an initial astrometric
solution. Subsequently, all detected objects contained in
the GSC were identified and used to obtain the final
solution, which was therefore in the GSC reference frame. The catalog
derived from FOCAS, which was used for the first spectroscopic run, had
equatorial positions derived from this solution.

A new astrometric solution was
carried out to verify whether imprecision or systematic effects in the
astrometry could have been a possible cause of the rather low success
rate in the spectroscopy during the 1992 run (see section 4 below).
The first step in this new solution involved defining a new primary
astrometric reference frame, which was done by measuring the positions
of a total of 67 stars from the ACRS catalog (Corbin \& Urban
1988), which has positions better than 0.1$''$, on
the Lick astrograph plate Ay-07697 (epoch 1974) using the Lick
Observatory Automatic Measuring Engine (AME, Klemola, Jones \& Hanson
1987). The second step was the redetermination of the right ascension and
declination on the astrographic plate of GSC stars that defined 
the secondary frame of reference. The solution for the secondary
reference frame used 442 stars and has $rms$ residuals 
of the order of 0.3$''$. A comparison 
between our measurements and those in the GSC is shown in Figure
1, where a systematic difference of up to 1$''$ is easily seen, and
the mean residuals between coordinates are shown in Table 1.
The ``corrected'' coordinates of GSC stars were then used to convert
the scan X and Y coordinates of FOCAS objects (referred to the origin
of the first scan) into equatorial coordinates.
The $rms$ residuals for this transformation are of the order of 0.5$''$,
and the total number of objects above detection limits in the
PDS-derived catalog is 13886.

Both plates were scanned by the Automatic Plate Measuring (APM) system
(Kibblewhite et al. 1984) 
at the Institute of Astronomy, University of Cambridge. The APM is a
high-speed microdensitometer that uses a scanning laser to measure the
plate transmission, and allows either detecting and classifying
objects in real-time when in catalog mode, or obtaining a digitized
image when in raster mode. The APM scans were carried out by M. Irwin
and N. Ellman in catalog mode, using a pixel size of 8 \micra
(0.54$''$ on the sky) and followed the standard procedure of
detection of objects. As described by Maddox et al. (1990) and Infante
\& Pritchet (1992), the data acquisition with the APM is made in two
steps. First, the plates are scanned to determine the local sky
background in each region of the plate, which is done by  dividing the
plate into a grid of 64 $\times$ 64 pixels; for each square of this
grid, the sky
background is determined by fitting the histogram of pixel values and
determining the value of the mode.  From the ensemble of these
squares the sky background at each pixel position is estimated 
by means of a bilinear interpolation (Maddox et al. 1990).

In the second pass objects are identified as groups of more than 12
connected pixels ($\sim$ 3.2 arc sec$^2$) with densities higher than
2 $\sigma$ above the sky background. This corresponds to a limiting
surface brightness of about 
25.5 \bj magnitudes arc sec$^{-2}$ and 24.5 \rf magnitudes arc
sec$^{-2}$ (Ellman 1994). The
limiting magnitude of these plates were estimated by Ellman to be 
\bj $\sim$ 23.5 and \rf $\sim$ 22.2. A series of parameters are
calculated for each object: the density-weighted X and Y centroid positions; 
the integrated  isophotal density, which corresponds to the APM
``magnitude''; the intensity-weighted second order moments (S$_{xx}$,
S$_{xy}$, S$_{yy}$) from which the size, shape and position angle of
objects may be derived; the peak density; and the number of pixels
above eight preset density threshold levels.
The equatorial coordinates are obtained from the positions of several 
PPM stars (R\"oser \& Bastian 1991) and the final
positions are usually accurate to about 0.5$''$ (Maddox et al. 1990).
These parameters are combined into a series of classifiers which allow
an objective separation between stars, galaxies, merged images, and
noise. The final catalog is then output with equatorial coordinates,
classifications, and the set of parameters described above. 

\subsection{Combination of Catalogs}

In this work we used APM data in a region slightly larger than our
survey, covering the whole R.A. range of the plates in the declination
range 29\deg\SP $\leq \delta \leq $ 30\deg\SP (Epoch 1950.0). The
total number of detections in each sub-catalog is 36619 for the F
plate and 36232 in J. Combining these catalogs it became possible to 
estimate the astrometric uncertainties, eliminate spurious detections,
and have an independent object classification and an estimate of the
color. Although small systematic differences in equatorial coordinates
dependent on the position on the plate were detected,
these are small and are always less than 0.3$''$ (Ellman 1994), mostly
in right ascension. This variation was taken into account through the
use of a linear fit when the combined catalog was made.

The master catalog used in this survey was defined from the J plate
scan, and although this choice was arbitrary, the smaller number of
objects down to the detection limit suggested that by using
it, the number of objects without colors might be minimized.
Besides the cut in declination, a further cut in apparent
magnitude, corresponding to approximately \bj= 21.5 was applied. This
was faint enough that no objects would be lost when defining the
redshift survey sample, yet bright enough that most of the objects were
real and would minimize the number of multiple matches for a same
object. No cut in magnitude was applied to the F plate catalog in
order to maximize the number of objects with colors. A
match was accepted whenever the separation (after correcting 
for the linear term described above) between
objects in both catalogs was less than 1.25$''$. This small
separation 
ensured that most objects would have only one match. Further searches
were made for non-matching objects at separations of 2$''$, 3$''$,
4$''$ and 5$''$, in order to include objects that are very bright or
that could be components of images classified as merged in one of the
APM catalogs. The combined catalog contains a total of over 13000
objects, of which about 10\% were classified in either or both of the
APM catalogs as merged.

This combined catalog was then correlated with that derived
from the PDS scans. There were three motivations for this: (1) it would
allow an independent verification of the astrometry; (2) it would
allow comparing independent classifications of objects, and (3)
it enabled us to separate merged images. One of the features of FOCAS
is that it allows
splitting merged images by making cuts at higher isophotal levels,
which improves the coordinates of objects at the possible expense of
magnitude accuracy. This of course depends on the relative apparent
magnitudes of the objects making up the merger. The total of
matching objects is 7710, and the comparison of coordinates resulted in
mean differences of the order of 0.3$''$ (see Table 1).
Of the 1162 objects classified as merged in the APM J catalog, 604 could be
split through this procedure. The rather small number of matches is
the result of a combination between the larger angular coverage of the APM
scans (5\deg) relative to the PDS scans, added to the fact 
that no magnitude cut was applied to the PDS catalog, which reaches
deeper than the adopted cut in \bj. 

The final catalog of objects after matching with the PDS scans
contains 13524 objects. The object class in this
catalog combines the classification from each of the scans
and attributes a score that ranges from high probability of being a
star or galaxy, when the
object presented the same classification in all catalogs, to
uncertain. This classification also took into account
cases in which an object was not contained in one of the scans of the F plate. All
objects presenting either conflicting classifications or that were
classified as merged and brighter than \bj= 20.5, were
inspected visually and classified using the images of PDS scans. 

Further checks on the astrometry of this combined catalog were
carried, and are also presented in Table 1. We have compared 
the APM coordinates with two unpublished catalogs
of the SA 57 region, the first coming from Kron (1980) which contains the
photometric corrections of Koo (1986), that we will refer to hereafter
as the KK catalog, and that of S. Majewski, which, although derived 
from KK, has a greater astrometric precision. From the comparison with
the catalogs in SA 57 we can see that our coordinates 
have an accuracy of about 0.4$''$. The differences we find with the KK
catalog have a systematic component, while the comparison with
Majewski's catalog shows only random errors. In addition to these
catalogs, which contain mainly faint objects, we compared our
positions with two surveys of bright galaxies, that of the drift scan
of Kent, Ramella \& Nonino (1993, hereafter KRN), and
the catalog of van Haarlem et al. (1993), which was also derived from
APM photometry. 
The comparison with the KRN catalog showed 96 galaxies in common.
By considering the values quoted by KRN for their positional
uncertainty (0.5 $''$ in R.A., 0.9 $''$ in decl.), our errors would be of the
order of 0.41 $''$ in R.A. and 0.87 $''$ in decl. The latter is much larger
than what is inferred from the comparison with the catalogs of fainter
galaxies.
If we assume that the APM coordinates are good to about 0.4$''$,
this result implies that the uncertainty in declination in KRN could be
underestimated. However, another possibility is that our measurement
of the centroids of bright galaxies could be affected by effects such
as saturation. This interpretation would also explain the rather large
dispersion in our comparison with the van Haarlem et al. (1993) sample,
although, in this case, we should stress that the number of objects in
common is very small  (19 galaxies).

\subsection{Photometry}

Although we had originally intended to calibrate the plates using CCD
photometry, which was obtained with the Mount Hamilton 1.0 m Nickel Telescope,
because of poor weather during our photometry runs,
only a very small number of calibrating objects were measured, and,
even for those, the quality of the data would not allow an adequate
calibration. Therefore, we have to rely on photographic data, and in
this calibration we have followed the procedure described by Ellman
(1994). 

As the catalogs produced both by APM and FOCAS
contain isophotal magnitudes, in order to obtain physically meaningful
colors, it is necessary to
measure ``total'' magnitudes for galaxies,
or to use the same metric aperture in all bands ($e.g.$, Bershady et
al. 1994). A previous attempt to
estimate total magnitudes for APM-generated catalogs was carried out
by Infante \& Pritchet (1992), who used the size information provided
by APM to divide galaxies into different surface-brightness classes
and then used this information to correct to total
magnitudes. However, as discussed
by Ellman (1994), this procedure could introduce discontinuities for
objects close to the surface brightness class limits. Therefore, in her
thesis, Ellman used a continuous correction that depended on the size
of an object relative to the size of a star with the same
magnitude. This correction was determined by fitting the stellar locus
of the size--magnitude diagram for each color with a low-order
polynomial, and then using this correction when solving the photometric
transformation.

To calculate the photometric transformation, we used the catalog of KK,
which contains total magnitudes measured in several bands.
Although this catalog is derived from photographic photometry, by using it
one can somewhat offset the lower precision of the photographic
photometry against the larger color corrections that would be
necessary if CCD photometry were used. In the calibration we used only
objects classified as galaxies 
in both APM catalogs and KK and that matched within 2$''$ of the KK
objects after the correction for small systematic offsets, calculated using the
catalog of Majewski. The total number of galaxies used in the
transformation was 115. These transformations were solved using
the IRAF PHOTCAL task:

\begin{equation}
b_{apm} = b_{1} + b_{2} {\rm b}_{J} + b_{3} ({\rm b}_{J} - {\rm
r}_{F}) + b_{4} Sc_{J} 
\end{equation}
\begin{equation}
r_{apm} = r_{1} + r_{2} {\rm r} _{F} + r_{3} ({\rm b}_{J} - {\rm r}_{F}) + r_{4} Sc_{F}
\end{equation}

\noindent {where we have followed  Ellman (1994) in considering a size
correction in the transformation (Sc$_J$ and Sc$_F$
respectively, where Sc=S$_{xx}$+S$_{yy}$).} As 
shown by Ellman (1994), the color term in the $b_{apm}$ transformation
(eq. 1) is negligible, and this transformation was solved setting
$b_3$ = 0. The $rms$ dispersion is $\sim$ 0.12 mag. By comparing
measurements of objects classified as galaxies in the KK sample, but
not necessarily in the APM catalogs, we find a mean difference of
\begin{equation}
{\rm b}_{J} - b_{KK} =   0.002 {\pm} 0.177
\end{equation}
\begin{equation}
{\rm r}_{F} - r_{KK} = -0.006 {\pm} 0.168
\end{equation}
\begin{equation}
({\rm b}_{J} - {\rm r}_{F}) - (b_{KK} - r_{KK}) = 0.009 {\pm} 0.149
\end{equation}
for 214 objects matching within 2$''$. A comparison between the photometry
in both catalogs is shown in the panels of Figure
2. Figure 2(a) shows the difference $b_{KK}$--\bj versus $b_{KK}$. For
magnitudes fainter than \bj = 19.0 the measurements are quite similar,
but at brighter levels, saturation causes our magnitudes to be fainter
than those in the KK catalog. 
Figure 2(b) shows the behavior of
color as a function of magnitude, and we can see that there is a tendency
of brighter objects to be redder in our 
sample, the expected behavior in the case of our measurements being
primarily affected by saturation in the J plate.

In order to have an independent check on the quality of our
photometry, in particular to search for the presence of systematic
effects dependent on position, we have compared our \rf photometry
measurements with those of KRN. The following transformation ($e.g.$,
Ellman 1994) was calculated: 
\begin{equation}
r_{KRN} =  r_{1} + r_{2} {\rm r}_{F} + r_{3} ({\rm b}_{J} - {\rm r}_{F}) + r_{4} \theta
\end{equation}
where $\theta$ is the distance from the plate center in arc minutes.
We find that the dependence on distance from plate center is very
small (0.004 $\pm$ 0.004 mag), a result which is of the same
order as found by Ellman (1994), who used independent CCD photometry
in the B and V bands. There is no significant improvement in the fit
by considering the color term. The final comparison kept both terms
constant and equal to zero, giving an $rms$ of 0.20 mag, with a
slight slope ($r_{2}$ = 1.14). 

From these comparisons, we estimate that the photometry errors are of
the order of 0.2 mag, which added in quadrature imply that the
errors of the color should be of the order of 0.3 mag. 
%
%

The final procedure that was carried out was to inspect the APM areal
profiles in order to estimate the number of saturated pixels for each
object. As shown by Ellman (1994), it is possible to estimate where
the emulsion becomes non-linear by
inspecting a diagram of peak intensity versus magnitude. By using the
same values as Ellman (1994), we find that the total number of galaxies
that have pixels above emulsion non-linearity in our catalog is
292. In Figure 3 we show the distribution of the number of galaxies in 
bins of 0.2 mag; Figures 3(a) and 3(b) show the logarithm of
total count of galaxies represented as solid lines, while the dotted
lines represent galaxies presenting saturated pixels in the J and F
plates respectively. For the blue plate, the 
number of galaxies presenting saturation is $\sim$ 80\% at \bj = 18.0,
while at \bj = 19.5 there are very few galaxies with pixels above
non-linearity, and for most of these, only 1 or 2 pixels are
saturated, so it is likely that the magnitudes 
are not very far off. For the F plate the corresponding limits are
16.5 for 80\% of galaxies with unreliable magnitudes, while at 18.0
there are no more galaxies with saturated pixels. We should note that
the worst cases for the F plate are galaxies presenting 4\% non-linear
pixels. 

\section{Spectroscopy}

The spectroscopic
observations were made with the KPNO Mayall 4.0 m 
telescope during two runs in 1992 February 27-29 and
1993 February 15-17. Because of bad weather on both runs, the total
observing time was equivalent to only 2 useful nights. The objects
were observed using the HYDRA multi-fiber positioner (Barden \etal
1992) and a bench-mounted spectrograph, with the KPNO T2KB chip.
This setup makes it possible to obtain
spectra of up to 97 objects simultaneously. We used
the red fiber cable, and the  wavelength ranged from 4100 to 
8100 \AA\SP with a dispersion of 2 \AA\SP per pixel and resolution of about 8
\AA\SP FWHM. Typically $\sim$ 75 
fibers were assigned to objects, while about 15 were assigned to
``empty'' sky positions.
The sky positions were selected by dividing the survey region into a
grid of 20$''$ squares and then counting the number of objects within
these squares using the full APM catalogs ($i.e.$, both catalogs without
matching and without any magnitude or object class cut). In this way,
we optimized the chance of not having the sky fibers contaminated by
faint background objects, as well as allowing a large choice of
possible sky positions; for a given field, there were typically about
twice as many sky positions as there were objects. The
fiber assignment was made though a code 
written by P. Massey which optimizes the number of fibers assigned to
objects taking into account the number of required sky positions, the
number of ``field orientation probes'' (that allow verifying the
astrometry and orientation of the field, but are also useful to check
the guiding or presence of clouds), and an optional weighting 
scheme, which 
is based on the rank of the object in a catalog. For our runs we
used the ``strong'' weighting scheme on a magnitude-ordered catalog.
During the second run, we could improve the fiber assignment through
the use of an optional feature that allowed the allocation of
fibers visually.
Integration times were usually 6000 s
divided into three exposures, in order to allow an efficient removal
of cosmic-ray events. Comparison exposures were taken before and after
each night to provide the wavelength calibration and to track shifts
in the positions of comparison lines.
During both runs we observed spectrophotometric and radial
velocity standard stars as well as a few bright galaxies,
which could serve as templates for the redshift determination,
and  to determine the radial velocity uncertainties and
zero points.
Data were reduced in the IRAF environment following standard
procedures ($e.g.$, Massey 1992) with removal 
of bias and dark current prior to the combination of images and
extraction of spectra. It was 
also found that removing scattered light improved the final results,
so that prior to the extraction of spectra, this correction was
done. In the 
IRAF tasks dealing with multi-fiber data, flat-fielding is usually
carried out during extraction of the spectra.  The
spectra were extracted using the DOHYDRA task written by
F. Valdes. The wavelength solution is also determined within this task,
and typically about 35 He-Ne-Ar lines were used in the final
transformations  with an $rms$ residual usually better than 0.10 \AA. 
The last step in this reduction is the sky subtraction which
used an average sky spectrum obtained combining the sky exposures of a
given setup. Redshifts were measured using the RVSAO package
(Kurtz et al. 1992) which measures redshifts using cross-correlation
and emission-line analyses. For
the cross-correlation analysis, a
series of templates was used, which included the standard CfA
templates distributed with RVSAO; composite spectra of stars and
galaxies measured for the Southern Sky Redshift Survey (da Costa et
al. 1989); high signal-to-noise ratio spectra of M31 (kindly provided by
C. Bellanger and V. de Lapparent) and NGC 7507; spectra of radial
velocity standard stars observed with HYDRA; and finally, composite
spectra of standard stars of Jacoby, Hunter, \& Christian (1984)
combined into different luminosity and spectral classes. This
assortment of templates permits optimizing the cross-correlation peak.
For all spectra we obtained, the final redshifts produced by RVSAO
were only accepted (or rejected) after visual inspection.
The internal errors for the redshift are estimated from the ratio
between the cross-correlation peak relative to the average noise
peaks, or from the dispersion between the radial velocities of each
emission line. When both are present, these errors are combined
($e.g.$, Tonry \& Davis 1979). We should stress that these
errors are certainly underestimated so that external estimates of the
uncertainties are required.

	In the panels of Figure 4, we show some spectra of objects
for which we could measure redshifts. The first five, Figures 4(a)
through 4(e), are typical representatives of objects for which we
could measure cross-correlation velocities, and the spectra are
ordered following decreasing redshift quality (see Section 4.1 below).
 Figures 4(f) through 4(j) show
the analogous case for objects presenting emission lines. None of
these spectra were corrected for the instrumental response. We show
for each object its identification, redshift, and redshift quality from
Table 3 and, in the case of Figure 4(e), from Table 5. We also show the
expected positions of some prominent absorption lines, which are used
in the visual confirmation of the redshift, as well as emission lines
that were either used for measuring the redshift or that would allow
a further confirmation on the correctness of the redshift. 

In general, the mininum signal-to-noise ratio (S/N) for which
cross-correlation velocities could be obtained was 3, the S/N being
measured in the interval between 6000 and 6200 \AA\SP, where the
spectra are generally flat and where there is no sky emission. In Fig.
5 we show the a plot of S/N $versus$ internal redshift uncertainty,
where we discriminate objects that produced both
cross-correlation and emission line velocities by using solid squares;
objects that produced cross correlation only are shown with open
squares, and objects that produced emission lines only are shown with 
crosses. We can see that emission line velocities could be measured
down to very low 
values of S/N, while very few objects presented cross-correlation 
velocities below S/N = 3; above S/N = 9, all objects presented
a cross-correlation velocity. We can see that the internal
errors are in their vast majority less than 100 kms$^{-1}$. In general,
the cross-correlation velocities obtained from different templates
usually agreed within the estimated internal errors.

To estimate our external uncertainty, we have compared our radial
velocities with measurements obtained by Ellman (1994),
van Haarlem et al. (1993), ZCAT (Huchra 1993) and the
radial velocity standard stars we observed. These measurements are 
presented in Table 2, and the column descriptions may be found in the
table notes. With ZCAT (Huchra 1993), there are 15 galaxies with
redshifts in common (not 13 as reported in paper I), with a mean
difference of 

$v_{here} - v_{ZCAT}$ = --3.7 $\pm$ 73.5 kms$^{-1}$

With the catalog of van Haarlem et al. (1993), there are 9 galaxies
for which we have also measured radial velocities; the mean difference
we find is 
 
$v_{here} - v_{vH}$ = --33.3 $\pm$ 66.2 kms$^{-1}$.

With the 5 galaxies in common with Ellman (1994) we find

$v_{here} - v_{Ellman}$ = --122.4 $\pm$ 41.1 kms$^{-1}$.

A 21 cm velocity has been obtained for one of our
faintest galaxies (M.A.G. Maia, private communication), which gives a
difference of $\sim$ --60 kms$^{-1}$. The radial velocity standard
stars were selected from the Astronomical Almanac (Blumberg \&
Boksenberg 1995) and we find for 6
measurements

$v_{here} - v_{AA}$ = --9.4 $\pm$ 13.6 kms$^{-1}$.

For all galaxies and stars, which make up a total of 35 objects, we
get 

$v_{here} - v_{others}$ = --30.2 $\pm$ 69.8 kms$^{-1}$.

This suggests that our external errors are probably less than 100
kms$^{-1}$, although there is a possibility that we might have a
zero-point shift of a few tens of kms$^{-1}$.


\section{Properties of the Redshift Catalog}
\subsection{Object Catalogs}

The total number of objects within the limits of this sample (from
$\sim$ 12$^h$ 47$^m$ to $\sim$ 13$^h$ 16$^m$, 29\deg\SP $\leq \delta
\leq$ 30\deg) is 13524, of which 7481 are stars, 4242 are galaxies,
137 are merged images and 1664 are noise. The latter were not removed $a\SP
priori$ from the catalog, because in a few cases these
corresponded to images of bright objects. For the observations, as well as
the analyses, we considered the region in the range 
12$^{h}$ 52$^{m}$ $\leq$ $\alpha$ $\leq$ 13$^{h}$ 12$^{m}$ and
29\deg\SP 05$'  \leq \delta \leq$ 29\deg\SP 54$'$ (epoch 1950.0). The range in
declination was chosen because it corresponds to the field size that could
be observed with HYDRA, while the R.A. range corresponds to the smallest
survey opening angle that could allow determining the presence of
wall-like structures at z $\sim$ 0.1; at this distance the opening
angle corresponds to $\sim$ 20 $h^{-1}$ Mpc.
At the nominal magnitude limit of the survey (\bj = 20), there are 1013
galaxies, 11 merged images and 3097 stars in the catalog.

%
The total number of redshifts we measured is 347, 326 of which are of
galaxies, 2 of which are quasars and 19 (5\%) of which are stars. During the
observations, objects fainter than the nominal limit were 
also considered as possible targets to place fibers, so that the
catalog contains redshifts of 47 galaxies fainter than \bj = 20.
The catalog of observed galaxies is presented in Table 3, and the columns
are described at the end of the table. 
 We should note also that for the quasars the redshifts were estimated
visually, as we were not successful in fitting these lines with the RVSAO 
software.
The estimate of the mean
surface brightness was calculated as  SB = \bj + 2.5 log$_{10}$ (area),
the area having been derived from the total number of pixels
contained by the outermost isophote detected by APM
transformed into square arc seconds. This expression is not entirely
correct, in the sense that 
the magnitude we are considering is a total magnitude and not the
isophotal magnitude measured within the limiting isophote used by
APM. However, we feel this should serve as an
adequate estimator of the surface brightness. The calculation of the
surface brightness within a circle of 2$''$ diameter, corresponding to
the approximate size of the fibers projected on the sky, while more
meaningful, would be affected by large uncertainties, as the catalog
output by the APM contains the measurement of the number of pixels
above 8 preset density levels. This means that it would 
be necessary to integrate the fitted image profile in order
to estimate the luminosity within 1$''$ radius. This estimate
would also have to take into account the characteristic curve as well as
make a correction for saturated pixels.

For completeness, 
in Table 4 we present a list with radial velocities and magnitudes of
objects in our catalog that turned out to be stars. In addition to
the 335 objects presented in Tables 3 and 4, we present in Table 5
a list of objects (8 galaxies and 4 stars) for which we
measured radial velocities, but which are not contained in our final
APM-derived catalog. These objects were measured during our first
run, but because of their faintness (or imprecise coordinates),
they have no matching object in
our APM-derived catalog. For these objects we have estimated r$_F$
magnitudes using a fit between the PDS instrumental magnitudes and 
calibrated \rf magnitudes for galaxies presenting both
measurements. None of the objects in Tables 4 and 5 will be  
considered further in the analyses that follow.

The number of galaxies with cross-correlation velocities is 155; those
whose redshift was estimated from emission lines number 126, while 39 had both
cross-correlation and emission line velocities. Thus, the proportion
of objects presenting emission lines is $\sim$ 52\%, which is similar
to that obtained by Bellanger et al. (1995) in their
deeper survey (R $\leq$ 20.5) close to the south Galactic pole. The
two quasars had 
been noted previously by Berger et al. (1991) as objects presenting
ultra-violet excess. 
 
In the analysis presented in paper I, in addition to the redshifts we
measured, we also included in our catalog measurements from ZCAT
(69 galaxies, some of which are unpublished, kindly
provided by J. Huchra); van Haarlem et al. (1993, 3 velocities); Ellman
(1994, 19 galaxies); Koo \& Kron (unpublished; 157 objects, of which
132 are galaxies) and one galaxy from  Boroson, Salzer, \& Trotter
(1993). 

%
%
\subsection{Catalog Completeness}
Two different effects play a role in the redshift completeness.
The first is the actual clustering of galaxies combined with the
limitations imposed by the use of multi-object spectroscopy. For
the mean density of objects at the limiting magnitude of
this project ($\sim$ 300 galaxies per square degree), it is
very difficult with the present technology to secure spectra for 100\% of
the objects in only one configuration. As mentioned above, the sample
contains 1013 galaxies with \bj $\leq$ 20, of which a total of 375
(37\%) were assigned to fibers. This then, represents the completeness
level imposed by the observational technique, though we should also
mention that this number also depends on the number of configured
fields (8), which was insufficient to cover the entire sample.

The second effect is the actual identification of features that would
allow measuring a redshift of objects that were observed.
Of the 375 observed galaxies  to b$_J$=20, 273 (73\%) yielded a redshift,
while 102 did not. An additional 165 objects fainter than the limiting
magnitude were observed without producing a measurable
spectrum. Thus, the efficiency, $i.e.$, our ability to 
obtain a spectrum that allowed measuring a redshift, is $\sim$ 70\%
at the limiting magnitude of the survey, while, when the total number of
positioned fibers is considered ($i.e.$, including objects fainter
than \bj = 20) $\sim$ 59\% yielded a redshift.

The total number of galaxies with redshifts to \bj = 20 is 383, which
includes in addition to our redshifts 110 measurements by other authors. 
The catalog completeness as a function of magnitude is presented in 
Figure 6(a) where the solid line represents the total fraction of galaxies
with redshift measurements and the dotted line represents the proportion
of galaxies contributed by this work. The  overall
completeness at the limiting magnitude is $\sim$ 35\%. This number is
slightly lower than that quoted in paper I ($\sim$ 40\%) because in
that work we used a preliminary calibration which did not make any
corrections of the APM isophotal magnitudes to obtain ``total''
magnitudes, nor did we take into account the dependence on color. When these
corrections are used, in particular the size correction, the magnitudes  
of galaxies become ``brighter'' and so more objects are included in
the sample. 

\subsection{Astrometry errors}

We have investigated some of the possible causes for this rather low
yield of redshifts for observed galaxies. The first we considered is
the existence of astrometrical errors (in particular for the 1992
run). In Figures 7(a) 
and 7(b) we show the distribution of residuals in R.A. and
decl. respectively, against R.A. from the match between the final combined
catalog with that used in the first run. We can see that in general the
residuals are less than 0.5$''$, but become fairly
dramatic in R.A. at the edges of the sample and in declination at
an R.A. $\sim$ 194\deg\SP (12$^h$ 56$^m$). From these figures, we
conclude that poor coordinates could indeed be a likely cause for the
low efficiency, particularly  at the eastern edge of the sample,
although we should note that this field was also the one presenting the
smallest number of configured objects (65), owing to the
lower projected surface density of objects.

\subsection{Surface Brightness}
Another parameter we considered is the mean surface
brightness. In Fig. 8 we present a histogram showing the distribution
in surface brightness for galaxies observed successfully (solid line),
while the dashed line represents galaxies whose spectra did not allow
measuring a redshift. Here we can see a significant drop at an SB
$\sim$ 24.3 mag arc sec$^{-2}$ for galaxies with measured
redshifts. The median surface brightness of galaxies with measured
redshifts (320) is 24.09 $\pm$ 0.27, compared to 24.39 $\pm$ 0.30 for
objects with unknown redshifts (182). Although the difference between
both values is only marginally larger than their estimated errors,
the figure suggests that the surface brightness is a limiting factor in
the successful measurement of spectra.

\subsection{Profile}
In Figure 9 we show a histogram of the profile slope, where, as above,
galaxies with measurable spectra are represented by solid lines, while
those which did not yield redshifts are represented by the dashed
line. The profile slope was derived fitting an 	exponential power-law
to the areal profiles of the APM catalog. The histogram 
suggests that the slope inclination could play a small role in the 
lack of measurable redshifts, in the sense that galaxies with flatter
slopes are somewhat less likely to yield redshifts, though
the difference between both distributions is really marginal.

\subsection{Color}
The galaxy color apparently is not an important parameter for the
successful observation of objects, as can be seen in Figure 10.
Although galaxies with redshifts peak at a bluer color, the median
values for the distribution are essentially the same
(\bj--\rf)$_{no\SP z}$  = 1.09 $\pm$ 0.26 and (\bj--\rf)$_{with\SP z}$ = 1.08
$\pm$ 0.24, respectively. As each of the observing runs used a
catalog derived from a different color, we verified whether this
could introduce a bias in the catalog of galaxies with redshifts (e.g.,
Ellman 1994).
For the first run we find (\bj--\rf) = 1.08 $\pm$ 0.25, while for the
second run it is (\bj--\rf) = 1.05 $\pm$ 0.25, which shows that there
is no difference between the samples. The overall median color of the
sample ($i.e.$, all galaxies, with and without measured spectra) is very
close to these values, being (\bj--\rf) = 1.02 $\pm$ 0.25.
Finally, we inspected objects presenting very extreme colors. In
general galaxies with very red colors either present saturation or
have images classified as merged by APM, while most of the very blue
objects are either faint, probably having large associated
uncertainties, or have perturbed appearances.

\section{Conclusion}

We have presented a catalog containing 328 redshifts and magnitudes of
galaxies observed in a 4\deg\SP $\times$ 0.67\deg\SP minislice close to the
direction of the north Galactic pole. A further 267 targets have been
observed but did not produce measurable spectra. The completeness
level of the catalog at its nominal limiting magnitude is $\sim$ 35\%.
The incompleteness in our data is the result of a combination of several
effects. The primary effect is the actual distribution of
galaxies combined with the limitations imposed by the
use of multiobject spectroscopy.  Thus, part of
the low redshift incompleteness is attributable to the insufficient
number of configurations,  as no field was
observed twice in two different configurations, while the clustering
of galaxies implies that some fields have more galaxies than others.
Our efficiency rate in obtaining a
redshift at the survey magnitude limit is $\sim$ 70\%.

We have shown in section 4 that other causes of incompleteness are
present, $e.g.$, poor coordinates
might have played a role in the low efficiency in redshift
acquisition, particularly for our easternmost field, but it
is clearly not the only cause. In our data, the mean surface
brightness  of objects seems to be an important limiting factor,
for which a break can be seen at $\sim$ SB = 24.3 \bj mag arc
sec$^{-2}$. Neither the color nor the profile shape seems to play an
important role in the ability of obtaining a measurable spectrum.
We have shown that many of the objects in our sample
presenting extreme colors have either saturated images in the J
plate or were classified as merged objects by the APM
software.

\acknowledgments
We would like to thank N. Reid for the loan of the Schmidt
plates; the director of NOAO for the use of the PDS; Ed Carder
for assistance during the PDS scans; the HYDRA team (S. Barden,
T. Armandroff, P. Massey, and L. Groves) for their help related to the
spectroscopy runs; A. Klemola for help in re-measuring the
astrometry; D. Mink for providing RVSAO;  S. Majewski for his
catalog, and the anonymous referee whose comments have helped to
improve the presentation of this paper. C.N.A.W. thanks S. Arnouts, L. da
Costa, and V. de Lapparent for discussions. We acknowledge the use of the NASA
Astrophysics Data System, the NASA National Space Science Data Center
and the NED/IPAC Extragalactic Database,
which is operated by the Jet Propulsion Laboratory, California
Institute of Technology, under contract with the National Aeronautics
and Space Administration. This work has been funded by the following
grants: CNPq 201036/90.8; NSF AST-8858203, AST-9023178; the
US-Hungarian Science and Technology Grant J.F. no. 010/90; California
Space Institute CS 6-89 and a Sigma Xi grant.

\clearpage
\makeatletter
\def\jnl@aj{AJ}
\ifx\revtex@jnl\jnl@aj\let\tablebreak=\nl\fi
\makeatother


\begin{deluxetable}{llrrcrccc}
\small
\tablewidth{0pc}
\tablecaption{Coordinate Residuals between Catalogs }
\tablehead{
\colhead{Catalog 1}              & \colhead{Catalog 2}             &
\colhead{N$_{\rm obj}$}          & \colhead{$\Delta\alpha$ ($''$)} &
\colhead{$\pm$}                  & \colhead{$\Delta\delta$ ($''$)} &
\colhead{$\pm$}                  & \colhead{$\Delta pos$ ($''$)}   &
\colhead{$\pm$} \nl
 \SP\SP\SP (1) & \SP\SP\SP (2) & (3) \SP\SP\SP & (4) \SP\SP\SP\SP &
  (5)  & (6) \SP\SP\SP\SP & (7)  & (8) & (9) \nl
}
\startdata
GSC       & AME        &   442 &  0.96 & 0.33 & -0.03 & 0.22 & 0.91 &
0.38 \nl 
PDS(1993) & PDS (1992) &   877 &  0.26 & 0.49 &  0.23 & 0.38 & 0.51 &
0.49 \nl
APM (J)   & APM (F)    & 25136 &  0.24 & 0.42 &  0.14 & 0.36 & 0.55 &
0.28 \nl
APM (J)   & PDS (1993) &  7710 &  0.03 & 0.25 &  0.01 & 0.24 & 0.27 &
0.22 \nl
KK        & APM (J)    &  1436 & -0.57 & 0.53 & -0.59 & 0.52 & 1.02 &
0.43 \nl
KK        & APM (F)    &  1393 & -0.26 & 0.54 & -0.52 & 0.50 & 0.84 &
0.40 \nl
Majewski  & APM (J)    &   815 &  0.12 & 0.29 & -0.01 & 0.34 & 0.38 &
0.26 \nl
KRN       & Final      &    96 &  1.22 & 0.65 & -1.28 & 1.26 & 2.08 &
0.87 \nl 
van Haarlem & Final    &    19 & -0.25 & 0.95 &  0.29 & 0.73 & 0.93 &
0.82 \nl
\tablecomments{Cols. (1) and (2) identification of catalog: GSC, Lasker \etal
(1990); AME, remeasurement of GSC objects using Lick Astrograph plate;
PDS, catalog derived from Focas scans, with coordinates derived 
from the Guide Star Catalog (PDS 1992) or from Lick Astrograph plate
(PDS 1993); APM, Catalogs derived from APM scans of J and F plates; KK,
coordinates from the catalog of D.C. Koo \& R.G. Kron (private communication)
for SA 57; Majewski, Catalog of S.R. Majewski (private
communication) for SA 57; KRN, Kent \etal 1993; van Haarlem: van
Haarlem \etal 1993; Final, final catalog of objects prepared for
this work. Col. (3) number of objects in common. Col. (4) mean
difference in the sense {\it catalog 1 -- catalog 2 \rm} for right
Ascension, Col. (5) its associated uncertainty. Col. (6) mean difference in
declination ({\it catalog 1 -- catalog 2 \rm}) and Col. (7)
uncertainty. Col. (8) mean quadratic difference between 
positions, combining the differences in right ascension and
declination and Col.(9) its associated uncertainty. }
\enddata
\end{deluxetable}

\clearpage

\makeatletter
\def\jnl@aj{AJ}
\ifx\revtex@jnl\jnl@aj\let\tablebreak=\nl\fi
\makeatother
\begin{deluxetable}{rlrrrrl}
\small
\tablewidth{0pc}
\tablecaption{Radial Velocities in common with other Sources}
\tablehead{
\colhead{NSER}                   & \colhead{Other Ident.} &
\colhead{v$_{here}$}             & \colhead{$\pm$}    &
\colhead{v$_{other}$}            & \colhead{$\pm$}    &
\colhead{Ref}\nl
{}  &  {} &  kms$^{-1}$ & kms$^{-1}$ & kms$^{-1}$ & kms$^{-1}$ & {}
\nl
(1) \SP\SP\SP & \SP\SP\SP\SP (2) & (3) \SP\SP\SP & (4) \SP\SP\SP &
(5) \SP\SP\SP & (6) \SP\SP\SP & (7) 
}
\startdata
  3160  & 12545+2912 &  8056 &  22 &   8019 & 100 & ZCAT \nl
  3251  &    vH   27 &  7511 &  18 &   7469 &  50 & vH   \nl 
  3251  & 12546+2919 &  7511 &  18 &   7516 & 100 & ZCAT \nl
  3318  &    vH   26 &  7516 &  23 &   7553 &  50 & vH   \nl
  3318  & 12548+2918 &  7516 &  23 &   7460 & 100 & ZCAT \nl
  3610  &    vH   25 &  7154 &  20 &   7071 & 300 & vH   \nl
  3819  & 12558+2913 &  7663 &  24 &   7560 & 100 & ZCAT \nl
  4261  & \nodata    &  7167 &  51 &   7227 &  20 & Maia \nl
  4938  & A1258+2913 &  6891 &  18 &   6995 &  67 & ZCAT \nl
  4975  &    vH   20 &  7313 &  28 &   7359 & 100 & vH   \nl
  4975  & I 842      &  7313 &  28 &   7275 &  15 & ZCAT \nl
  5207  & A1258+2944 & 25151 &  30 &  25192 & 100 & ZCAT \nl
  5465  & I 843      &  7497 &  29 &   7387 &  25 & ZCAT \nl
  5539  & MK 61      & 17211 &  16 &  17121 & \nodata & ZCAT \nl
  5561  &    vH   18 &  7047 &  28 &   7097 & 100 & vH   \nl
  5561  & I4088      &  7047 &  28 &   7099 &   6 & ZCAT \nl
  5573  &    vH   19 &  6390 &  31 &   6490 &  50 & vH   \nl
  5763  &    vH   15 &  7307 &  21 &   7362 & 100 & vH   \nl 
  5763  & 12597+2931 &  7307 &  21 &   7412 & 100 & ZCAT \nl
  5764  &    vH   16 &  8056 &  23 &   8127 & 100 & vH   \nl
  6607  & 1301+2929  & 50006 &  94 &  49997 & \nodata & ZCAT \nl
  6854  &    vH   14 &  6825 &  19 &   6873 &  50 & vH   \nl
  7366  &    B2  533 &  7078 &  24 &   7175 &  24 & E94  \nl 
  7366  & 13030+2934 &  7078 &  24 &   7158 & 100 & ZCAT \nl
  7646  &    B2  513 & 15271 &   8 &  15353 &  22 & E94  \nl 
  7766  &    B2  521 &  7389 &  11 &   7532 &  62 & E94  \nl 
  7915  &    B2  512 & 25256 &  11 &  25362 &  96 & E94  \nl 
  8081  &    B2  509 & 54491 &   9 &  54675 & \nodata & E94  \nl 
  9848  & N 5004A    &  7217 &  28 &   7262 &  20 & ZCAT \nl
\nodata & N 4472     &   931 &  21 &    997 &  10 & ZCAT \nl
\nodata & HD  66141  &    53 &   5 &     71 &   1 & AA   \nl
\nodata & HD 107328  &    35 &  22 &     36 &   1 & AA   \nl
\nodata & HD 123782  &   -26 &  14 &    -13 &   1 & AA   \nl
\nodata & HD 132737  &   -15 &   2 &    -24 &   1 & AA   \nl
\nodata & HD 132737  &   -49 &   4 &    -24 &   1 & AA   \nl
\tablecomments{Col. (1) serial number in our main catalog. Col. (2) other
identification. Col.(3) radial velocity measured in this
work and in Col. (4) its associated internal uncertainty. Col. (5)
radial velocity, Col. (6), uncertainty and Col. (7) reference of
other measurements.}
\tablerefs{Blumberg \& Boksenberg (1995, AA); Ellman (1994, E94);
M.A.G. Maia (private communication, Maia) ; van Haarlem et al. (1993, vH); 
Huchra (Private Communication, ZCAT)}
\enddata
\end{deluxetable}

\clearpage
\setcounter{page}{32}

\makeatletter
\def\jnl@aj{AJ}
\ifx\revtex@jnl\jnl@aj\let\tablebreak=\nl\fi
\makeatother

\begin{deluxetable}{rccrrrrr}
\small
\tablenum{4}
\tablewidth{0pc}
\tablecaption{Redshifts and Magnitudes for Observed Stars}
\tablehead{
\colhead{NSER}      & \colhead{R.A.} &
\colhead{Dec.}      & \colhead{v}    &
\colhead{$\pm$} & \colhead{q}    &
\colhead{b$_J$}     & \colhead{b$_J$--r$_F$} \nl
{} & 2000.0 & 2000.0 & kms$^{-1}$ & kms$^{-1}$ & {} & {} & {}\nl
(1) &  (2)  &  (3)   &  (4) \SP \SP & (5) \SP \SP & (6) & (7) & (8)
}
\startdata
 2407 & 12:55:56.98 & 29:14:27.2 &    30 &  43 & 2 & 18.79 & 1.03 \nl 
 2607 & 12:55:21.46 & 29:27:24.6 &   -13 &  41 & 5 & 18.87 & 1.02 \nl 
 3237 & 12:57:35.65 & 29:21:07.4 &   -48 &  93 & 5 & 19.99 & 1.59 \nl 
 3449 & 12:57:03.78 & 29:25:23.9 &  -120 &  85 & 5 & 19.60 & 1.59 \nl 
 3526 & 12:57:13.72 & 29:12:43.1 &    33 &  64 & 5 & 19.45 & 1.52 \nl 
 3535 & 12:57:14.70 & 29:08:36.5 &   -71 &  39 & 4 & 18.59 & 0.91 \nl 
 4389 & 12:59:02.85 & 29:11:20.4 &   105 &  97 & 4 & 19.92 & 1.55 \nl 
 5467 & 13:01:09.90 & 29:03:40.3 &   101 &  78 & 4 & 20.22 & 1.64 \nl 
 5772 & 13:02:41.11 & 29:01:55.9 &    -7 &  62 & 5 & 19.73 & 1.44 \nl 
 5965 & 13:02:02.13 & 29:21:15.3 &   222 &  75 & 4 & 18.59 & 1.08 \nl 
 6068 & 13:02:15.72 & 29:12:33.8 &   -35 &  35 & 5 & 19.68 & 1.25 \nl 
 6627 & 13:03:25.98 & 29:30:25.0 &   -39 &  41 & 5 & 18.81 & 1.53 \nl 
 6729 & 13:04:38.43 & 29:22:07.3 &  -235 & 128 & 4 & 19.33 & 0.68 \nl 
10009 & 13:11:00.55 & 29:34:04.4 &   -56 & 118 & 4 & 20.45 & 1.67 \nl 
10741 & 13:12:35.79 & 29:14:55.9 &    66 &  59 & 3 & 20.46 & 1.77 \nl 
\tablecomments{Col. (1) serial number in our main catalog. Col. (2) right
ascension. Col. (3) declination. Col. (4) radial velocity. Col. (5)
internal error of radial velocity. Col. (6) redshift quality derived
from the number of identified features in spectrum. Col. (7)
magnitude. Col. (8) color.}
\enddata
\end{deluxetable}

\clearpage

\makeatletter
\def\jnl@aj{AJ}
\ifx\revtex@jnl\jnl@aj\let\tablebreak=\nl\fi
\makeatother

\begin{deluxetable}{rccrrrrr}
\small
\tablenum{5}
\tablewidth{0pc}
\tablecaption{Redshifts of Objects not in the Main Catalog}
\tablehead{
\colhead{PDS Id.}                & \colhead{R.A.} &
\colhead{Dec.}                   & \colhead{z}    &
\colhead{$\pm$ } & \colhead{t}    &
\colhead{q}                      & \colhead{estimated r$_F$}\nl
{} & 2000.0 & 2000.0 & {} & kms$^{-1}$ & {} & {} & {}\nl
(1) &  (2)  &  (3)   &  (4) \SP \SP & (5) \SP \SP & (6) & (7) & (8)
}

\startdata
  882  &   12:55:50.71  & 29:11:22.49 &  0.19788 &  54 & x & 1 & 19.84  \cr
  686  &   12:57:14.41  & 29:12:24.55 &  0.35489 &  36 & e & 5 & 19.23  \cr
  698  &   12:58:46.52  & 28:58:26.20 & -0.00062 & 237 & x & 3 & 19.01  \cr
  889  &   12:59:02.57  & 29:14:22.19 & -0.00050 & 107 & x & 5 & 19.51  \cr
  704  &   13:01:16.24  & 29:34:03.58 &  0.32904 & 105 & x & 3 & 19.29  \cr
  638  &   13:02:46.72  & 29:11:47.81 &  0.39613 &  62 & c & 4 & 19.13  \cr
  655  &   13:03:57.29  & 29:13:48.69 &  0.38763 & 104 & x & 2 & 19.15  \cr
  959  &   13:04:04.18  & 29:16:22.42 & -0.00014 &  44 & x & 3 & 19.89  \cr
  630  &   13:04:48.69  & 29:10:34.71 &  0.38778 & 143 & x & 2 & 19.11  \cr
  988  &   13:11:06.28  & 29:14:38.22 &  0.00008 & 125 & x & 2 & 20.33  \cr
  978  &   13:11:32.38  & 29:21:35.50 &  0.23847 &  95 & x & 1 & 20.64  \cr
  858  &   13:12:46.08  & 28:58:13.23 &  0.32588 &  60 & x & 1 & 19.74  \cr
\tablecomments{Col. (1) serial number in catalog derived from PDS
scans used in the first observing run. Col. (2) right
ascension. Col. (3) declination. Col. (4) 
observed redshift. Col. (5) internal radial velocity error. Col. (6)
type of redshift: cross-correlation (x), emission (e) or combined
cross-correlation and emission (c). Col. (7) redshift quality derived
from the number of identified features in the spectrum. Col. (8) r$_F$
magnitude estimated from a fit between the instrumental magnitude
measured with the PDS microdensitometer calibrated r$_F$
magnitudes. The latter have uncertainties of $\sim$ 0.20 mag}
\enddata
\end{deluxetable}

\clearpage


\begin{figure}
\caption{Residuals between Lick Astrograph and GSC coordinates of GSC stars
within the boundaries of this survey. The length of the arrow at the top
left corner corresponds to 1 arc second. The systematic difference
between coordinates of both solutions is immediately apparent.}
\end{figure}


\begin{figure}
\caption{Comparison between magnitudes calculated using the
transformations described in section 2.2 compared to the Koo and
Kron catalog. Filled circles represent galaxies used in the
calibration, while those that were not used are shown as open triangles.
These comparisons are shown as (a) a difference in magnitudes in b$_J$,
and (b) differences in color against b$_J$. }
\end{figure}


\begin{figure}
\caption{Distribution in magnitudes of galaxies with pixels above
emulsion non-linearity in each band. These are represented by short
dashed lines in (a) and (b), while the total number of
galaxies is shown as a solid histogram.}
\end{figure}


\begin{figure}
\caption{Examples of galaxy spectra measured for this work. (a$\sim$e)
Typical examples of galaxies for which we were
successful in measuring cross-correlation redshifts; these
figures are ordered in decreasing redshift quality, which is related to
the number of features that can be identified in the
spectra. (f$\sim$j) Spectra of galaxies for which we secured
emission-line redshifts. At the top left corner we show the galaxy
identification, redshift and its quality from Table 3, except
in the case of (e), where the data are taken from Table 5. Under
the spectra, the expected positions of some prominent absorption lines
are noted. Emission lines are identified above the spectra.}
\end{figure}


\begin{figure}
\caption{Distribution of signal-to-noise ratio against internal radial velocity
uncertainty. Filled squares represent galaxies with both emission and
cross-correlation velocities; open squares stand for galaxies with
cross-corrrelation only, and crosses represent those with emission only. The
ability of measuring radial velocities at low signal-to-noise levels for
objects with emission lines is immediately apparent. In general, the
smallest signal-to-noise ratio for which cross-correlation velocities
can be measured is about 3, while above 9 all objects have measured
cross-correlations.}
\end{figure}


\begin{figure}
\caption{Redshift completeness as a function of apparent
magnitude, for the whole sample of galaxies. The dashed line
represents the fraction of galaxies that were contributed by this work.}
\end{figure}


\begin{figure}
\caption{Residuals of coordinate differences between the final catalog and
the catalog in the first observing run plotted as function of (a) right
ascension and (b) declination. This plot suggests that poor
coordinates could have been a cause of the rather low detection rate
of objects during the first observing run.}
\end{figure}


\begin{figure}
\caption{Histogram showing the mean surface brightness distribution
for galaxies with redshifts (plotted as solid lines) and for galaxies
that were observed, but for which no features could be identified in
the spectrum (dashed line). There is a significant break in the
success rate for galaxies fainter than about 23.4 mag arc sec$^{-2}$.}
\end{figure}


\begin{figure}
\caption{Distribution of profile slope measured in the J plate 
for galaxies with measured redshifts (solid line) and unsuccessfully
observed (dashed line). Both distributions are similar, suggesting
that the profile shape is not an important parameter in determining
whether a redshift will be successfully obtained.}
\end{figure}


\begin{figure}
\caption{Color distribution of galaxies with measured redshifts
(solid line) and without (dashed line). This diagram suggests that our
ability of obtaining a measurable spectrum in not correlated with
the galaxy color.}
\end{figure}

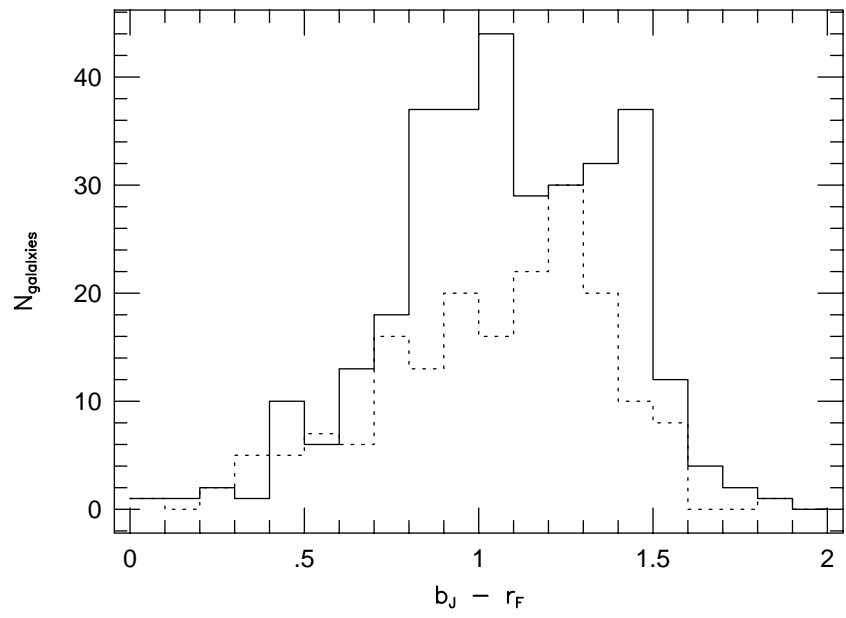
\end{document}